\documentclass{ifacconf}
\usepackage{graphicx}
\usepackage{epsfig}
\usepackage{amsmath}
\usepackage{amssymb}
\usepackage{natbib}        

\renewcommand{\Im}{\mathrm{Im}}
\renewcommand{\Re}{\mathrm{Re}}

\renewcommand{\imath}{\mathrm{i}}

\usepackage{color} 

\begin{document}
\begin{frontmatter}

\title{Transient behavior in systems with time-delayed feedback}

\author[TU]{Robert C. Hinz}%
\author[TU]{, Philipp H{\"o}vel}
\author[TU]{, and Eckehard Sch{\"o}ll}

\address[TU]{Institut f{\"u}r Theoretische Physik, Technische
Universit{\"a}t Berlin, 10623 Berlin, Germany (e-mail: schoell@physik.tu-berlin.de)}
 

\begin{abstract}
We investigate the transient times for the onset of control of steady states by time-delayed feedback. The optimization
of control by minimising the transient time before control becomes effective is discussed analytically and numerically,
and the competing influences of local and global features are elaborated.
We derive an algebraic scaling of the transient time and confirm our findings by numerical simulations in dependence on
feedback gain and time delay.
\end{abstract}

\end{frontmatter}

\section{Introduction}
\label{sec:intro}
Control of nonlinear dynamical systems has attracted much attention since the seminal ground-breaking work of Ott,
Grebogi, and Yorke \citep{OTT90}, and a large variety of different control schemes have been proposed
\citep{FRA07,SCH07}.
One particularly successful method called \textit{time-delayed feedback control} was introduced by Pyragas in order to
stabilize periodic orbits embedded in a strange attractor of a chaotic system \citep{PYR92}. In this scheme, the
difference between the current control signal $s(t)$, which is generated from some system variables, and its
time-delayed counterpart $s(t-\tau)$ yields a control force that is fed back to the system. If the
time delay matches the period of the target orbit, this control force vanishes. Thus, time-delayed feedback is a
noninvasive control method.
The Pyragas scheme has been successfully applied for the control of both unstable steady states \citep{HOE05,DAH07} and
periodic orbits in different areas of research ranging from mechanical 
\citep{BLY08,SIE08} and neurobiological systems
\citep{SCH08,SCH09c} to optics
\citep{TRO06,SCH06a,FLU07,DAH08b} 
as well as semiconductor devices
\citep{SCH00,SCH09}
and chemical systems \citep{BAL06}. 
On the theoretical part, previous
work includes investigations on analytical properties \citep{JUS04,AMA05}, asymptotic scaling for large time delays
\citep{YAN06,YAN09}, and limitations of this powerful control method \citep{FIE07,JUS07}. However, the
transient dynamics of time-delayed feedback control has not been investigated systematically.

In this paper, we aim to obtain a deeper insight into the control mechanism and its efficiency by an analysis of
transient times before control is achieved. The transient times and their scaling behavior have been studied in
particular in the context of chaotic transients \citep{TEL08}, e.g., in spatially extended systems, where
supertransients were found.
Here, we consider the case of steady states  in linear and nonlinear systems which are subject
to time-delayed feedback control. We investigate a generic system beyond a supercritical Hopf bifurcation.
Thus, the unstable fixed point can be treated as a linearized normal form above the bifurcation. Furtheron, we focus
also on global aspects \citep{LOE04,HOE07a}.

This paper is organized as follows: In Sec.~\ref{sec:uss}, we investigate the transient times of control of an unstable
focus in the presence of time-delayed feedback within a linear model and relate this quantity to the eigenvalues of
the controlled system. Section~\ref{sec:upo} is devoted to effects of time-delayed feedback on transient times of fixed
point control in a nonlinear system under the influence of stable periodic orbits. We finish with a conclusion
in Sec.~\ref{sec:conclusion}.

\section{Linear transients}
\label{sec:uss}
In this Section, we consider an unstable fixed point of focus type which is subject to time-delayed feedback
\citep{HOE05,DAH07}. It can be described within a generic model in center-manifold coordinates by a linear system
which corresponds to the normal form close to, but above a supercritical Hopf bifurcation whose nonlinear effects will
be discussed in Sec.~\ref{sec:upo}. The dynamic equations are given by
\begin{subequations}\label{eq:uss}
\begin{eqnarray}
\dot x(t)&=& \lambda x(t) + \omega y(t) - K \left[x(t)-x(t-\tau)\right]\\
\dot y(t)&=& -\omega x(t) + \lambda y(t)  - K \left[y(t)-y(t-\tau)\right],
\end{eqnarray}
\end{subequations}
where $\lambda>0$ corresponds to the regime of unstable fixed point and $\omega\neq 0$ is the intrinsic
frequency of the focus. The control parameters $K\in\mathbb{R}$ and $\tau$ $\in\mathbb{R}$ denote the feedback gain and
time delay,
respectively. In complex notation $z=x\pm i y$, Eqs.~(\ref{eq:uss}) become
\begin{eqnarray}\label{eq:uss_polar}
\dot z(t)&=& (\lambda\pm i\omega) z(t) - K \left[z(t)-z(t-\tau)\right].
\end{eqnarray}
Similarly, using $z=r e^{i\varphi}$ with amplitude $r\geq 0$ and phase $\varphi$, Eq.~(\ref{eq:uss_polar}) can be
rewritten in the uncontrolled case ($K=0$) as
\begin{eqnarray}\label{eq:uss_polar2}
\dot r(t) = \lambda r(t),\qquad \dot \varphi(t) =\omega.
\end{eqnarray}
The amplitude equation will serve as starting point of our analytical derivations presented later in this Section.

The stability of the system (\ref{eq:uss}) can be inferred from the characteristic equation
\begin{eqnarray}\label{eq:uss_chareq}
\left[\Lambda +K\left( 1-e^{-\Lambda \tau}\right) -\lambda \right]^2 + \omega^2 &=&0,
\end{eqnarray}
where the fixed point is stable if the real part of all eigenvalues $\Lambda\in\mathbb{C}$ is negative. Note that
solutions of this transcendental equation can be found analytically using the multi-valued Lambert function $W$ which is
defined as the inverse function of $f(z)=ze^{z}$ for $z\in\mathbb{C}$ \citep{HOE05}:
\begin{eqnarray}\label{eq:uss_Lambert} 
 \Lambda&=& \frac{1}{\tau}W\left(K\tau e^{-(\lambda\pm i \omega)\tau+K\tau}\right)+\lambda\pm i \omega-K.
\end{eqnarray}

\begin{figure}[t]
\centering
\includegraphics[width=\linewidth]{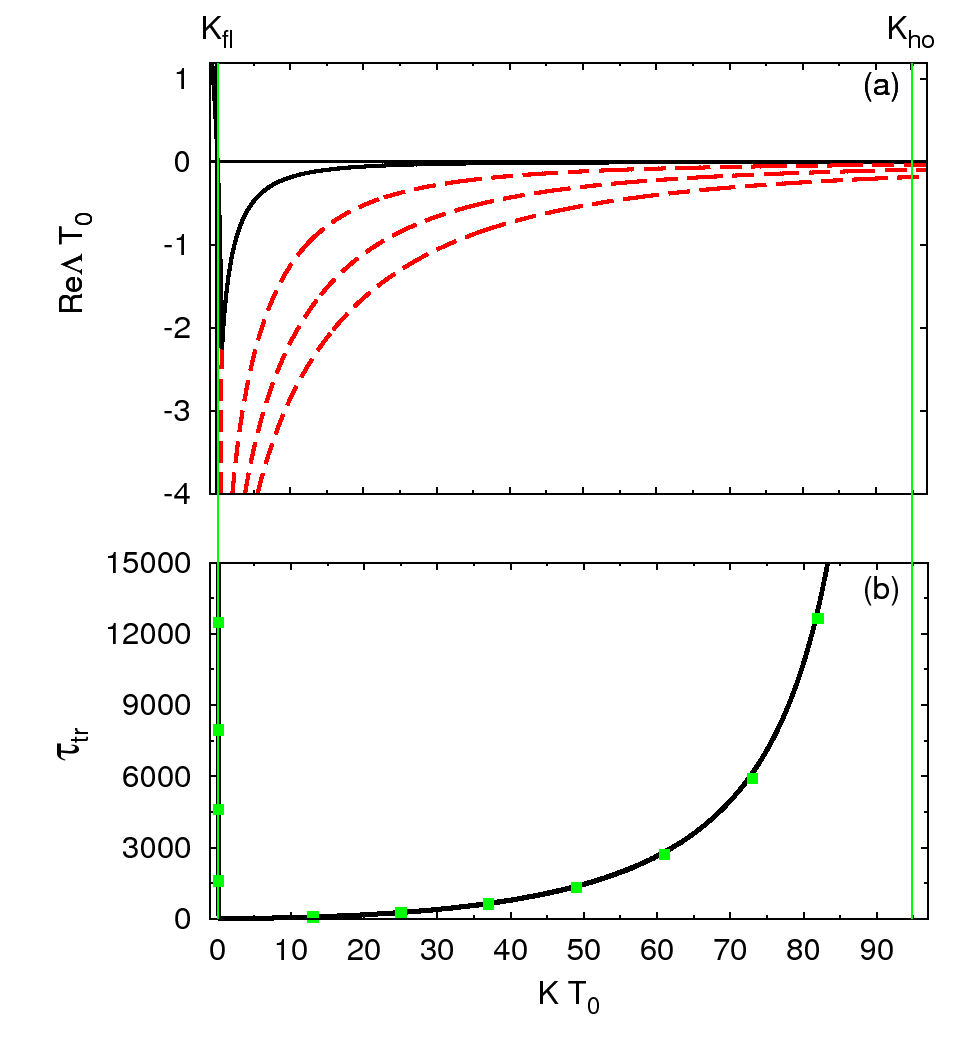}
\caption{(a): Largest real part of the complex eigenvalues $\Lambda$ vs $K$ for a fixed time delay
$\tau=T_0/2=1$ (solid curve). The dashed red curves show additional modes. (b): Transient time $\tau_{tr}$ in
dependence on the feedback gain $K$. The solid curve corresponds to the analytical formula~(\ref{eq:uss_transtime2}) and
the green dots refer to values obtained by numerical simulations of Eqs.~(\ref{eq:uss}). The green
lines at $KT_0=\lambda T_0/2=0.1$ and $KT_0=94.76$ correspond to the flip ($K_{fl}$) and Hopf threshold ($K_{ho}$),
respectively. Parameters: $\lambda=0.1$, $\omega=\pi$, $\tau=T_0/2=1$, $r_0=0.1$, and
$\epsilon=0.001$.}
\label{fig:focus_evttime}
\end{figure}

In the following the initial conditions are taken from the uncontrolled system for $t\in[-\tau,0)$ and the control is
switched on at $t=0$. Figure~\ref{fig:focus_evttime}(a) depicts the largest real parts of the eigenvalues $\Lambda$
calculated from the
characteristic equation~(\ref{eq:uss_chareq}) in dependence on the feedback gain $K$. The time delay is fixed at
$\tau=T_0/2=1$ with the intrinsic timescale $T_0=2 \pi/\omega$. The dashed red curves refer to lower eigenvalues
arising from $-\infty$ in the limit of vanishing $K$. In the absence of a control force, the system is unstable with
$\Re\Lambda=\Re\Lambda_0=\lambda>0$. As the feedback gain increases, the largest real part becomes
smaller and changes sign at $K_{fl}=\lambda/2$ where the system gains stability in a flip bifurcation. Above this change
of stability, the largest real part collides with a control-induced branch and forms a complex conjugate pair. For even
larger values of $K$, the system becomes unstable again in a Hopf bifurcation at $K_{ho}$. Both threshold values of $K$
are marked as green vertical lines.

In the following, we will derive an analytical relation between the solutions of the characteristic equation and the
transient times. Starting from an initial distance $r_0$, the transient time $\tau_{tr}$ to reach a neighborhood
$\epsilon\ll r_0$ around the fixed point is given for the uncontrolled case of Eq.~(\ref{eq:uss_polar2}) by
the following expression
\begin{eqnarray}\label{eq:uss_transtime}
\tau_{tr}(r_0) &=&   \int^{\epsilon}_{r_0} \frac{dr}{\lambda r}= -\frac{1}{\lambda} \log \left(
\frac{r_0}{\epsilon}\right).
\end{eqnarray}
Note that $\lambda$ corresponds to the real part of the uncontrolled eigenvalue $\Lambda_0$. Time-delayed feedback
influences the eigenvalues according to the characteristic equation~(\ref{eq:uss_chareq}) such that the transient time
in the presence of the control scheme becomes 
\begin{eqnarray}\label{eq:uss_transtime2}
\tau_{tr}(r_0) &=& -\frac{1}{\Re\Lambda} \log \left(\frac{r_0}{\epsilon}\right),
\end{eqnarray}
where $\Re\Lambda$ denotes the largest real part of the eigenvalues which is depicted by the black solid curve in
Fig.~\ref{fig:focus_evttime}(a).

\begin{figure}[t]
  \centering
   \includegraphics[width=0.7\linewidth]{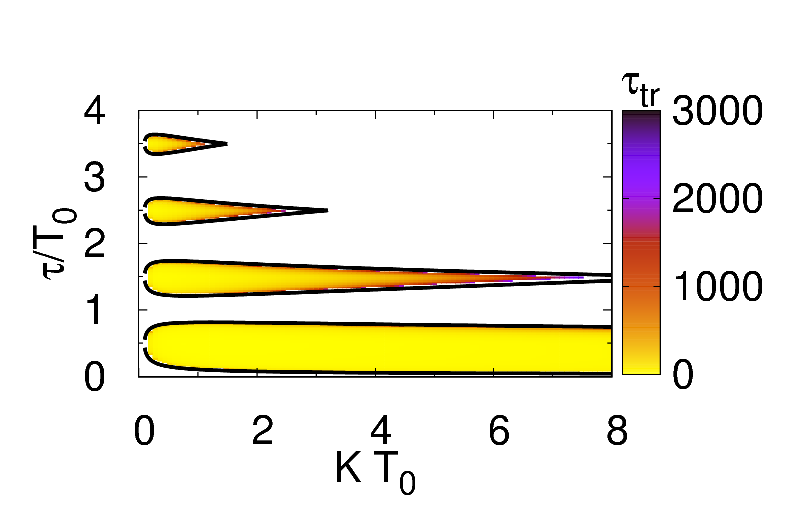}
   \caption{Transient times for the unstable focus in the $(K,\tau)$ plane as color code.
The black curves show the boundary of the domain of control, i.e., $\Re\Lambda=0$, obtained from the
characteristic equation~(\ref{eq:uss_chareq}). Parameters: $\lambda=0.1$, $\omega=\pi$, $r_0=0.38$, and
$\epsilon=0.01$.}
   \label{fig:focus_Ktauplane}
\end{figure}

Figure~\ref{fig:focus_evttime}(b) displays the transient time in dependence on the feedback gain. The solid curve
corresponds to the transient time calculated from Eq.~(\ref{eq:uss_transtime2}). The green dots depict the
transient time $\tau_{tr}$ obtained from numerical simulations of the system's equation~(\ref{eq:uss_polar}), where
$\tau_{tr}$ is measured as the duration to enter a neighborhood of radius $\epsilon=0.001$ starting from an initial
distance $r_0=0.1$. One can see that the transient time diverges at the flip bifurcation and the Hopf threshold where
the largest real part becomes zero. This is indicated by the solid green lines at $K_{fl}$ and $K_{ho}$. There
is a broad optimum of transient times in a wide range of feedback gain $K$. Thus the efficiency of control is not very
sensitive to the choice of $K$, which is not evident from mere inspection of the eigenvalues
(Fig.~\ref{fig:focus_evttime}(a)).

Figure~\ref{fig:focus_Ktauplane} shows the transient times in the plane parame\-trized by both control
parameters $K$ and $\tau$ as color code. Note that there are islands of stability separated by areas for
which the control fails to stabilize the fixed point. Similar to Fig.~\ref{fig:focus_evttime}(b), the value of the
transient time becomes arbitrarily large at the boundary of control. The black solid curves correspond to the 
boundary of control, i.e., vanishing real part of the largest eigenvalue in Eq.~(\ref{eq:uss_chareq}). Optimum control,
i.e., minimum transient times, occurs in the center of the tongues of stability $\tau=(2n+1) T_0/2$, $n\in
\mathbb{N}_0$.

Since the eigenvalues $\Lambda$ are given by the Lambert function in Eq.~(\ref{eq:uss_Lambert}),
the transient time $\tau_{tr}$ cannot be written in terms of elementary functions. Nevertheless, an approximation can be
given near the values of $K$ where the largest real part changes its sign, i.e., at the boundaries of stability. Using a
linear approximation of the dependence of $\Re\Lambda$ on the feedback gain $K$ at the flip and Hopf bifurcation
points $K_{fl}$ and $K_{ho}$, one finds for the leading eigenvalue
\begin{equation}\label{eq:f_series}
\Re\Lambda[K]\approx\Re\Lambda'\left[K_{fl/ho}\right] \left(K-K_{fl/ho}\right)
\end{equation} 
with $\Lambda'\left[K_{fl/ho}\right]=d\Lambda/dK|_{K=K_{fl/ho}}$. Writing the eigenvalue $\Lambda$ in terms of the
Lambert function $W$ as given by Eq.~(\ref{eq:uss_Lambert}) this derivative becomes
\begin{subequations}
\begin{align}
\Re\Lambda'[K] &= \frac{d}{dK}\left[ \frac{W ( \kappa )}{\tau} + \lambda \pm i \omega - K
\right] \\
		&= -1 + \frac{1}{\tau}\frac{d}{d\kappa} W(\kappa) \; \frac{d\kappa}{dK}.
\label{eq:f_seriesfirstorder}
\end{align} 
\end{subequations}
with the abbreviation $\kappa=K \tau e^{-(\lambda \pm i \omega) \tau + K \tau}$. Using the analytical expression for
the derivative of $W$ \citep{COR96}
\begin{equation}
\frac{d}{d\kappa}W(\kappa)=\frac{W(\kappa)}{\kappa [1+W(\kappa)]}
\end{equation} 
for $\kappa \ne 0$, and using Eq.~(\ref{eq:uss_Lambert}) to express $W(\kappa)$ in terms of $\Lambda$, it
follows from Eq.~(\ref{eq:f_seriesfirstorder})
\begin{align}
\Re\Lambda' =& \frac{\left[(\Im\Lambda-\omega)^2+\lambda  (\lambda -K)\right]
\tau -\lambda}{K \left[(\Im\Lambda-\omega) ^2+(K-\lambda )^2\right] \tau ^2+2 K (K-\lambda ) \tau +K} 
\label{eq:f_seriesfirstorder_eval}
\end{align}
Therefore, for the expression around the flip threshold, where $K=K_{fl}=\lambda/2$ and $\Im\Lambda=\omega$
holds, one obtains 
\begin{eqnarray}\label{eq:uss_real_flip}
\Re\Lambda[K] &\approx& \frac{4}{\lambda\tau - 2 } \left(K-K_{fl}\right)
\end{eqnarray} 
and for the transient time
\begin{equation}\label{eq:uss_trans_flip}
\tau_{tr}(K)\approx  -\frac{\lambda\tau - 2 }{4} \left(K-K_{fl}\right)^{-1}
\log\left(\frac{r_0}{\epsilon}\right).
\end{equation} 
In analogy, one finds near the Hopf threshold at $K_{ho}$, where the imaginary part of the largest
eigenvalue is given by $\Im\Lambda=\omega \pm \sqrt{(2 K_{ho} -\lambda)\lambda}$ \citep{YAN06}, the following
expression
\begin{equation}\label{eq:uss_real_hopf}
\Re\Lambda[K]\approx \frac{[\lambda  \left(\tau K_{ho}-1\right)] \left(K-K_{ho}\right)}{\tau ^2 K_{ho}^3+2
\left(K_{ho}-\lambda\right) \tau K_{ho}+K_{ho}},
\end{equation}
which yields
\begin{equation}\label{eq:uss_trans_hopf}
\tau_{tr}(K)\approx \frac{\tau ^2 K_{ho}^3+2 \left(K_{ho}-\lambda\right) \tau K_{ho}+K_{ho}}{[\lambda\left(\tau
K_{ho}-1\right)] \left(K-K_{ho}\right)}\log\left(\frac{r_0}{\epsilon}\right).
\end{equation} 
Hence, in both case a power-law scaling of the transient time $\tau_{tr}\sim\left(K-K_{fl/ho}\right)^{-1}$ is obtained.

We note that a supertransient scaling
\citep{TEL08}
of the form $\tau_{tr}(K)\sim
\exp\left[C\left(K-K_{fl/ho}\right)\right]^{-\chi}$ with positive constants $C$ and $\chi$ cannot be found because the
derivatives of $\Re\Lambda$ at $K=K_{fl/ho}$ do not vanish
\begin{equation}
\lim\limits_{K \rightarrow K_{fl/ho}}{\frac{d }{d K}}\Re\Lambda\neq0,
\end{equation} as can be seen in Fig.~\ref{fig:focus_evttime}(a).

\begin{figure}[t]
   \centering
   \includegraphics[width=\linewidth]{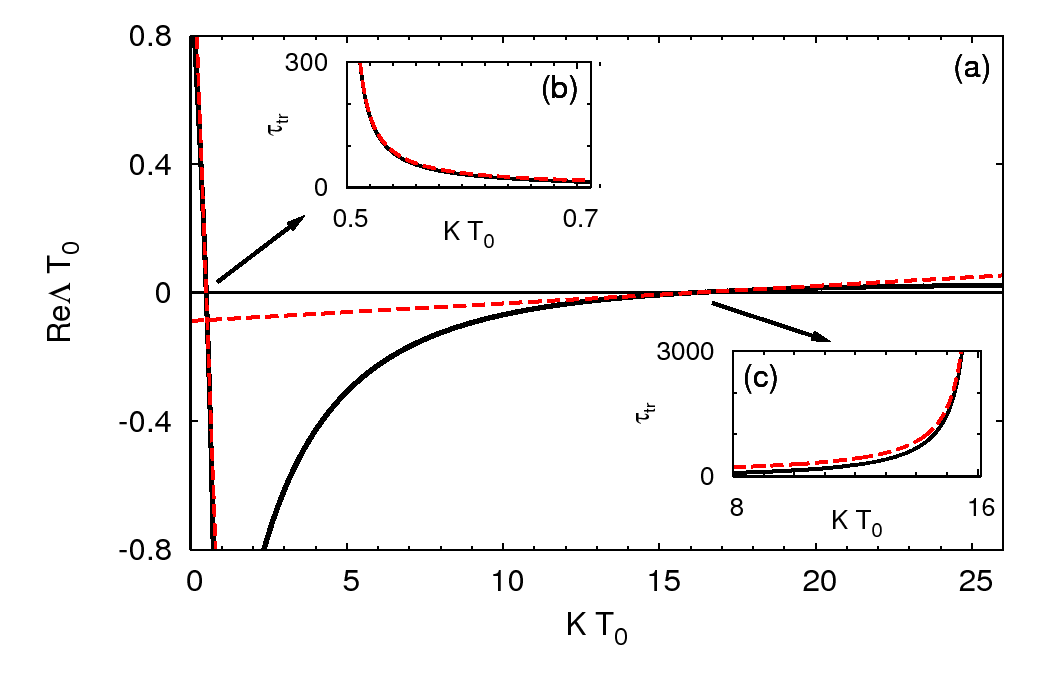}
   \caption{Linear approximation for $\Re\Lambda$ at $K=K_{fl}$ and $K=K_{ho}$ as dashed red lines.
The black curve shows the largest real part as in Fig.~\ref{fig:focus_evttime}(a). The insets (b) and (c) show the
transient times at the flip and Hopf threshold according to Eqs.~(\ref{eq:uss_trans_flip}) and
(\ref{eq:uss_trans_hopf}), respectively. Parameters as in Fig.~\ref{fig:focus_evttime}.}
\label{fig:focus_trans_linear}
\end{figure}

Figure~\ref{fig:focus_trans_linear} depicts the linear approximations of $\Re\Lambda$ at the flip and Hopf
threshold
points $K_{fl}$
and $K_{ho}$ as dashed red lines according to Eqs.~(\ref{eq:uss_real_flip}) and (\ref{eq:uss_real_hopf}),
respectively. The insets (b) and (c) display the transient time around these $K$-values as given by
Eqs.~(\ref{eq:uss_trans_flip}) and (\ref{eq:uss_trans_hopf}), respectively. While the linearization yields a good
approximation at the flip threshold, its deviations are more pronounced at the Hopf threshold because $\Re\Lambda$
changes here at a slower rate.


\section{Nonlinear transients}
\label{sec:upo}
This Section is devoted to investigations in \textit{nonlinear} systems containing periodic orbits. We will consider the
Hopf normal form as a generic model
given by the following equation (Stuart-Landau oscillator)
\begin{equation}\label{eq:hopf}
\dot z(t) = (\lambda + i\omega)z(t)+(a+ib)|z(t)|^2 z(t).
\end{equation} 
As an extension to Eq.~(\ref{eq:uss_polar}) of Section~\ref{sec:uss}, a cubic nonlinearity is taken into account
with real coefficients $a$ and $b$. Before addressing the effects of time-delayed feedback on this system, we will
briefly review the uncontrolled case in terms of the transient times. Using again amplitude and phase variables, i.e., $z=re^{i\varphi}$, Eq.~(\ref{eq:hopf})
becomes \begin{subequations}
\begin{eqnarray}\label{eq:hopf_normrad1} 
\dot r(t)&=& \left[\lambda + a r(t)^2 \right] r(t) \\
\label{eq:hopf_normrad2}
\dot \varphi(t) &=& \omega  + b r(t)^2.
\end{eqnarray}
\end{subequations}
Note that the equation for the amplitude~(\ref{eq:hopf_normrad1}) yields a periodic orbit with
$r_{PO}=\sqrt{-\lambda/a}$. In the following, we will restrict our consideration to the case $a<0$ which corresponds to
a supercritical Hopf bifurcation, i.e., there exists a stable periodic orbit for $\lambda>0$.

Similar to Sec.~\ref{sec:uss}, the transient time can be calculated from the amplitude
equation~(\ref{eq:hopf_normrad1}) as follows
\begin{subequations}
\begin{align}\label{eq:T_hopf_I}
\tau_{tr}(r_0)=& \int^{r_f}_{r_0} \frac{dr}{r(\lambda + a r^2)}\\
\label{eq:transient_time} 
	      =& - \frac{1}{\lambda} \log \left( \frac{r_0}{r_f} \right) + \frac{1}{2\lambda} \log \left(
\frac{r_0^2-r_{PO}^2}{r_f^2-r_{PO}^2}\right),
\end{align}
\end{subequations}
where $r_0$ denotes an initial amplitude and the final amplitude $r_f$ is chosen as $\epsilon$ or $r_{PO}\pm\epsilon$
for the analysis of the transient time concerning the fixed point at the origin and the periodic orbit, respectively.
Note that the coefficients in front of the two logarithmic functions correspond to the inverse of the real part of the
eigenvalue $\lambda$ of the fixed point and the Floquet exponent $\Lambda_{PO}=-2\lambda$ of the
supercritical periodic orbit, respectively.

\begin{figure}[t]
  \centering
  \includegraphics[width=0.9\linewidth]{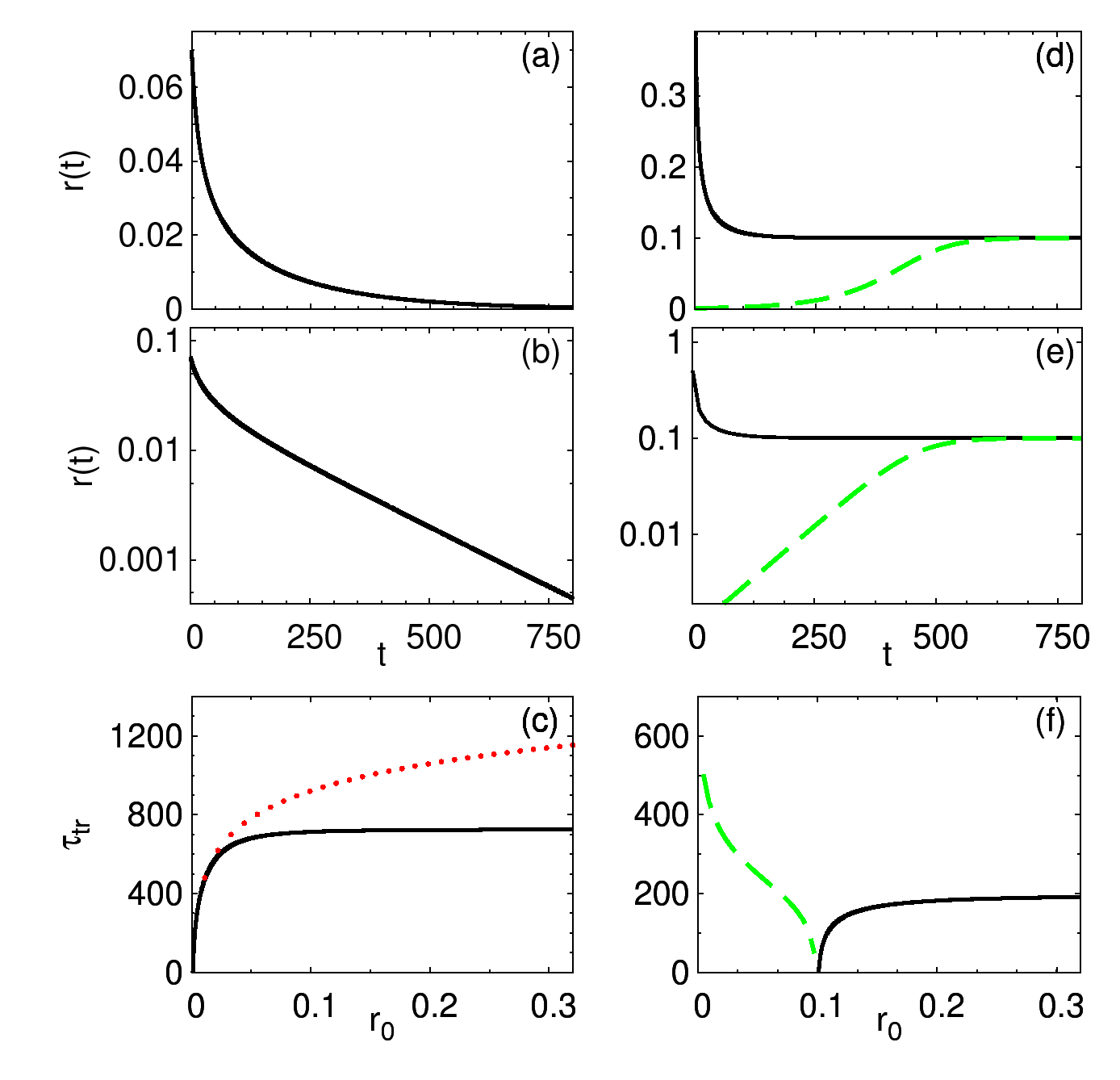}
  \caption{Time series of the amplitude $r$ (top: linear, middle: logarithmic scale) and transient time
(bottom) for a system with supercritical Hopf bifurcation (a)-(c): $\lambda=-0.005$ and (d)-(f): $\lambda=0.01$; dashed
green and solid curves denote an initial radius smaller and greater than the radius of the stable periodic orbit
$r_{PO}=\sqrt{-\lambda/a}$. (c) and (f) display the transient times for $\epsilon=0.001$. The dotted red curve in panel
(c) refers to the linear case of Sec.~\ref{sec:uss}. Parameters: $\omega=\pi$, $a=-0.1$, and $b=1.5$.}
   \label{fig:hsuper_anal2}
\end{figure}

Figure~\ref{fig:hsuper_anal2} depicts the time series of the amplitude $r=|z|$ and the transient time $\tau_{tr}$ of
the Hopf normal form~(\ref{eq:hopf}), where panels (a)-(c) and (d)-(f) correspond to a parameter value below
($\lambda=-0.005$) and above ($\lambda=0.01$) the Hopf bifurcation, respectively. Note that the time series is displayed
in linear as well
as logarithmic scale. 

Below the bifurcation, the stable fixed point at the origin is the only invariant solution. Panel (b) shows that this
fixed point is approached exponentially. Panel (c) depicts the transient time in dependence on the initial amplitude
$r_0$ according to Eq.~(\ref{eq:T_hopf_I}). The dashed (red) curve refers to the linear case discussed on
Sec.~\ref{sec:uss} showing the difference to the nonlinear system Eq.~(\ref{eq:hopf}). 

Above the bifurcation (Fig.~\ref{fig:hsuper_anal2}(d)-(f)), the fixed point is unstable and the trajectory approaches
the periodic orbit $r_{PO}=\sqrt{-\lambda/a}$. Note that the dotted (green) and solid curves correspond to initial
conditions $r_0$ inside and outside this periodic orbit, respectively. For initial conditions close to the origin the
transient time becomes arbitrarily large as the trajectory needs more time to leave the vicinity of the repelling fixed
point.

Next, we consider effects of time-delayed feedback control. Applying
this control scheme to the Hopf normal form, the system's equation~(\ref{eq:hopf}) becomes
\begin{align}\label{eq:hopf_tdas}
  \dot z(t) = \left[\lambda + i\omega+(a+ib)|z(t)|^2\right] z(t)-K[z(t)-z(t-\tau)]
\end{align}
with the feedback gain $K\in\mathbb{R}$ and time delay $\tau$. In the following, we will keep the time delay fixed at
$\tau=1=T_0/2$, as in Sec.~\ref{sec:uss}, but set initial conditions as $x=r_0, y=0$ for $t\in[-\tau,0)$.

\begin{figure}[t]
  \centering
  \includegraphics[width=0.7\linewidth]{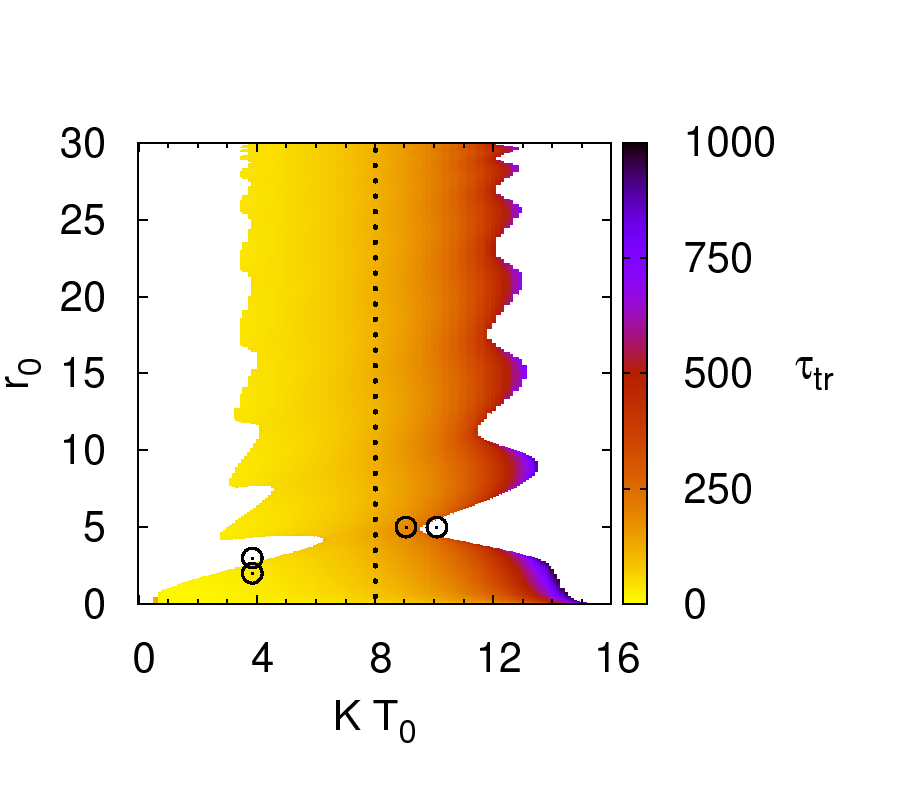}
  \caption{Transient times to reach the fixed point at the origin in the $\left(K,r_0\right)$ plane as
color code. The dotted line at $KT_0=8$ corresponds to the feedback gain used in
Fig.~\ref{fig:hsuper_induziert_a-0.1_K4}. The white areas indicate parameter pairs for which the trajectory
does not reach the fixed point. The circles mark parameters used in Fig.~\ref{fig:hsuper_induziert_a-0.1_tseries}.
Parameters: $\lambda=0.5$, $\epsilon=0.001$, $a=-0.1$, $b=1.5$, $\omega=\pi$, $\tau=1$.}
   \label{fig:hsuper_induziert2}
\end{figure}

Figure~\ref{fig:hsuper_induziert2} depicts the transient time $\tau_{tr}$ in dependence on the feedback gain $K$ and
initial amplitude $r_0$ as color code. The delay time $\tau=T_0/2$ was demonstrated to be an optimal choice
in the purely linear system discussed in Sec.~\ref{sec:uss}. The white areas correspond to parameter values for
which the trajectory does not reach the fixed point. For $K=0$ the fixed point is unstable. For a certain finite
non-zero feedback gain, however, the fixed point can be stabilized by time-delayed feedback. The transient
time $\tau_{tr}$ becomes larger as $K$ increases even further until control is lost again similar to
Fig.~\ref{fig:focus_evttime}(b).


%

\begin{figure}[t]
  \centering
  \includegraphics[width=\linewidth]{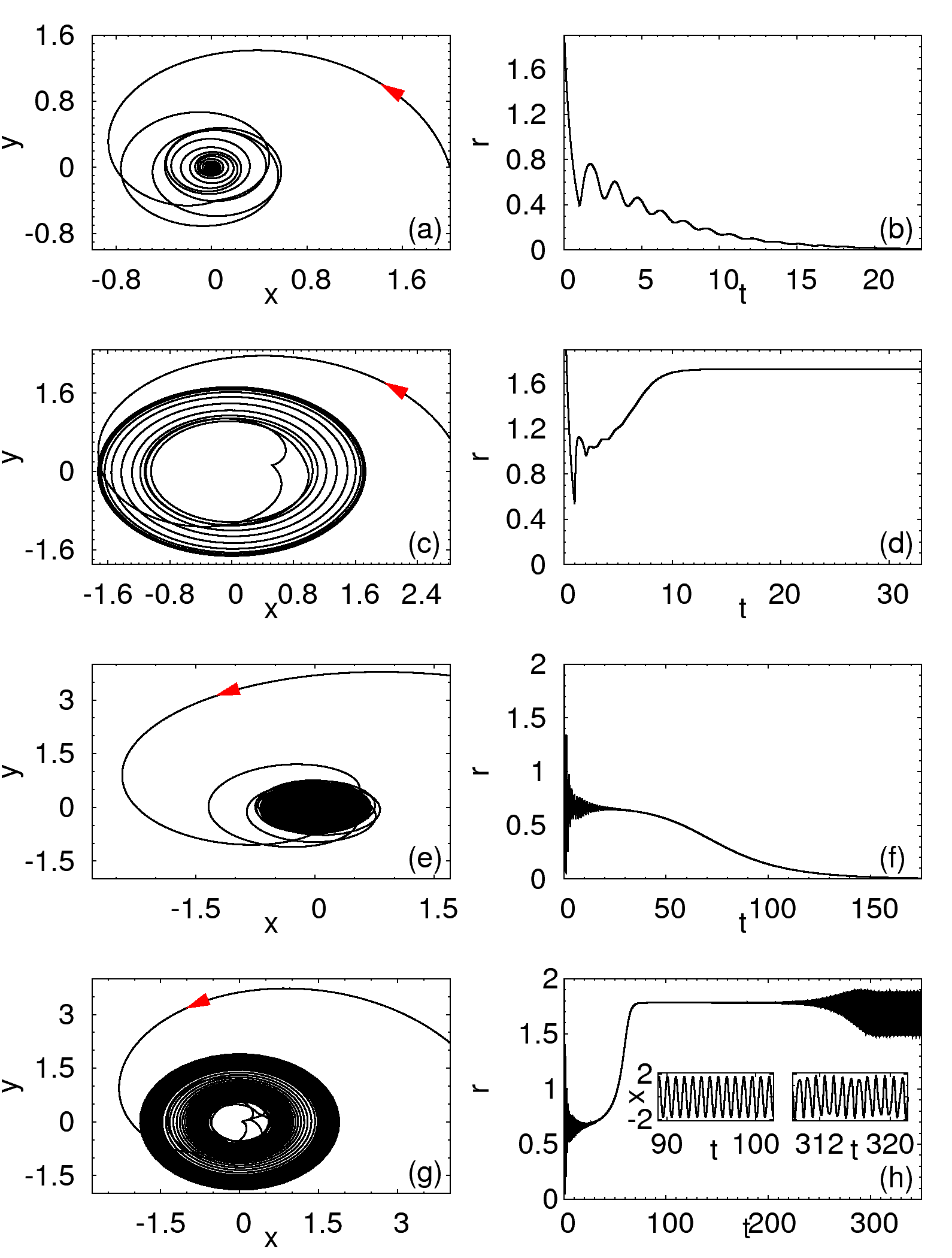}
  \caption{Phase portraits (left column) and time series $r(t)=|z(t)|$ (right column) for different
combinations of $r_0$ and $K$. The red arrows indicate the direction of the trajectory. (a),(b) $KT_0=4,
r_0=2$; (c),(d) $KT_0=4, r_0=3$; (e),(f) $KT_0=9, r_0=5$; (g),(h) $KT_0=10, r_0=5$.
These values of $r_0$ and $K$ are marked in Fig.~\ref{fig:hsuper_induziert2} as circles.
Other parameters as in Fig.~\ref{fig:hsuper_induziert2}.}
   \label{fig:hsuper_induziert_a-0.1_tseries}
\end{figure}

For a better understanding of the success and failure of the time-delayed feedback scheme,
Fig.~\ref{fig:hsuper_induziert_a-0.1_tseries} shows the time series for selected combinations of the feedback gain $K$
and
initial amplitude $r_0$. Panels (a),(c),(e), and (g) depict the trajectory in the $(x,y)$ phase space where the arrow
indicates the direction. Panels (b),(d),(f), and (h) display the time series of the amplitude $r=|z|$. The
parameters $K$ and $r_0$ are chosen as follows: Figures~\ref{fig:hsuper_induziert_a-0.1_tseries}(a)-(d) illustrate the
behavior at the left boundary of the yellow region of Fig.~\ref{fig:hsuper_induziert2} with fixed feedback
gain $KT_0=4$. While the fixed point is still stabilized in panels (a),(b) for $r_0=2$, the control fails for
$r_0=3$ in panels (c),(d) where a delay-induced stable periodic orbit is asymptotically reached. On the right boundary
of Fig.~\ref{fig:hsuper_induziert2} and for fixed $r_0=5$, panels (e),(f) correspond again to successful stabilization
of the steady state at the origin for $KT_0=9$, whereas slightly larger feedback gain ($KT_0=10$) of panels (g),(h)
results asymptotically in a delay-induced torus (see inset in Fig.~\ref{fig:hsuper_induziert_a-0.1_tseries}(h)). This
explains the modulation of the stability range in Fig.~\ref{fig:hsuper_induziert2} due to resonances with delay-induced
periodic or quasi-periodic orbits which reduce the basin of attraction of the fixed point.

\begin{figure}[t]
  \centering
  \includegraphics[width=0.7\linewidth]{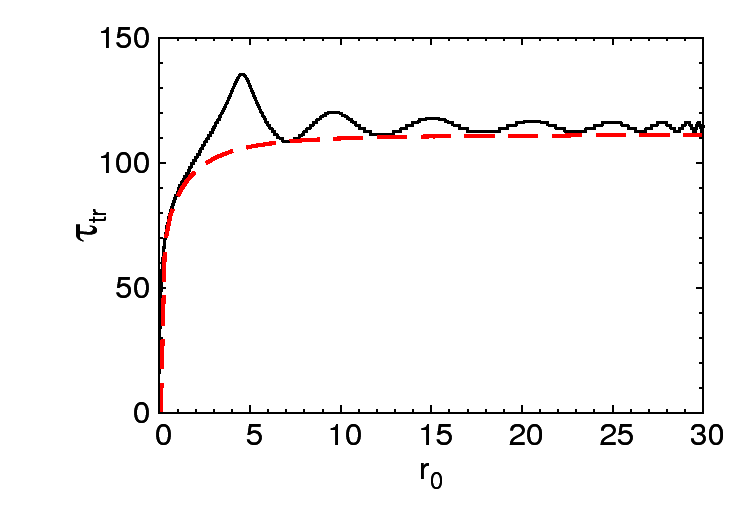}
  \caption{Transient time $\tau_{tr}$ in dependence on the initial radius $r_0$ for a fixed feedback gain
$KT_0=8$. The dashed red curve refers to the analytical formula of the uncontrolled case given by
Eq.~(\ref{eq:transient_time}). Parameters as in Fig.~\ref{fig:hsuper_induziert2}.
}
   \label{fig:hsuper_induziert_a-0.1_K4}
\end{figure}

In contrast to the linear case, the controllability displays an interesting non-monotic dependence upon the initial
condition $r_0$, with a strongly reduced range of control at certain values of $r_0$ resembling resonance-like
behavior. Although the fixed point is locally stable under time-delayed feedback control in the whole range of $K$
shown in Fig.~\ref{fig:focus_evttime}(a), the global behavior is strongly modified by a finite size of the basin of
attraction. This is reflected in the effect of the initial condition upon the transient time as displayed in
Fig.~\ref{fig:hsuper_induziert_a-0.1_K4} for fixed feedback gain $KT_0=8$, i.e., along the vertical dotted line
indicated in Fig.~\ref{fig:hsuper_induziert2}. One can see a strong increase of $\tau_{tr}$ for small $r_0$ which is
followed by a damped oscillatory behavior. For large initial amplitudes, this curve approaches the value corresponding
to the uncontrolled system which is added as dashed red curve and calculated from Eq.~(\ref{eq:transient_time})
for a real part $\lambda=-0.062$ of the eigenvalue of the stabilized fixed point and $a=-0.003$ (fitted parameters).
The strong modulation of the transient time with $r_0$ can be explained by the following qualitative argument. It
follows from Fig.~\ref{fig:focus_Ktauplane} that control works best in the linear system if $\tau=T_0(2n+1)/2$ where
$T_0=2\pi/\omega$ is the intrinsic timescale of the uncontrolled system, and it fails for $\tau=nT_0$,
$n\in\mathbb{N}$. Now, in the nonlinear system, the effective angular velocity $\omega^*$ changes with distance $r_0$
from the fixed point according to $\omega^*=\omega+br_0^2$ by Eq.~(\ref{eq:hopf_normrad2}). For $b>0$ the angular
velocity increases with increasing initial radius $r_0$. Hence, the period $T^*=2\pi/\omega^*$ decreases with increasing
radius.  Thus, for fixed $\tau=T_0/2$ and increasing $r_0$, the ratio $\tau/T^*=(1+br_0^2/\omega)/2$ successively passes
alternating half-integer and integer values. This suggests that the resonance conditions for best and worst control are
alternatingly satisfied, even though the fixed point itself is still linearly stable. By this we can explain the
modulation of the transient time in Fig.~\ref{fig:hsuper_induziert_a-0.1_K4}. This is, of course, a simplified argument,
since it does not take into account that not only the angular velocity is shifted by the nonlinearity, but also the
radial velocity changes nonlinearly with radius by Eq.~(\ref{eq:hopf_normrad1}). 

\begin{figure}[t]
  \centering
  \includegraphics[width=0.7\linewidth]{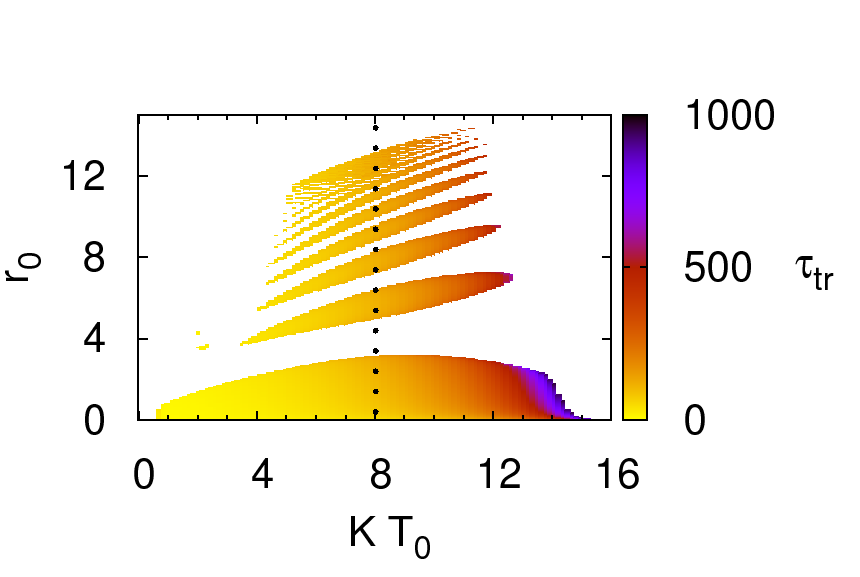}
  \caption{Same as Fig.~\ref{fig:hsuper_induziert2} for $a=0$.}
   \label{fig:hsuper_induziert_a0}
\end{figure}

In order to clearly separate the effects of modified angular and radial velocity, we will now consider the limit case
$a=0$ where the nonlinearity affects only the angular velocity $\omega^*$ and hence $T^*$. 
Figure~\ref{fig:hsuper_induziert_a0} illustrates the behavior of $\tau_{tr}$ for $a=0$ in the $(K,r_{0})$-plane. In
contrast to Fig.~\ref{fig:hsuper_induziert2}, the domain of control is no longer connected but consists of several
islands of stability. They are separated by white regions where control fails. The sequence of these white regions with
increasing $r_0$ can be qualitatively explained by the condition $\tau=nT^*$, $n\in\mathbb{N}$, i.e.,
$r_0\sim\sqrt{n-1/2}$. Here, the transients do not converge to the fixed point although this is linearly stable,
but rather to a delay-induced orbit. This is clear indication of the finite basin of attraction of the fixed
point, and of complex global effects in the nonlinear system. Note that our simple qualitative explanation does not
describe the exact position of the gaps of stabilization, since $\omega^*$ changes with increasing time, and the
condition $\tau=nT^*$ holds only for the linear system. Within the stability islands, larger feedback gain $K$ leads
to longer transients.

\begin{figure}[t]
  \centering
  \includegraphics[width=0.7\linewidth]{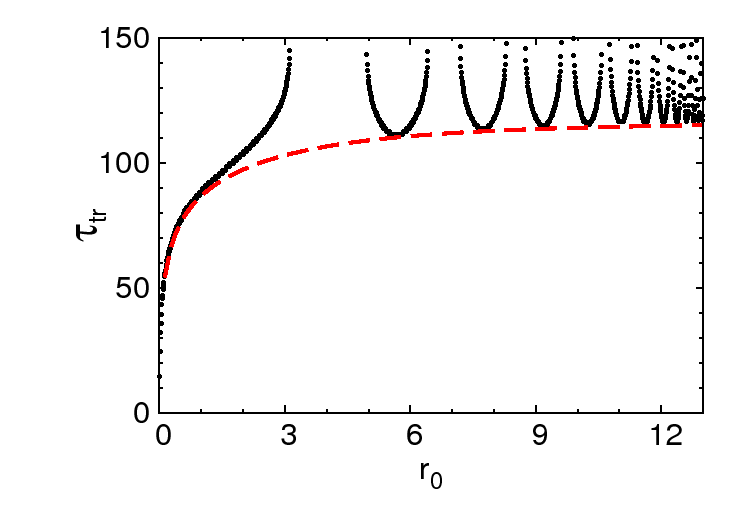}
  \caption{Same as Fig.~\ref{fig:hsuper_induziert_a-0.1_K4} for $a=0$.}
\label{fig:hsuper_induziert_a0_K4}
\end{figure}

Similar  to Fig.~\ref{fig:hsuper_induziert_a-0.1_K4} , Fig.~\ref{fig:hsuper_induziert_a0_K4} shows a vertical cut of 
Fig.~\ref{fig:hsuper_induziert_a0} for fixed feedback gain $KT_0=8$  as indicated by the dotted line. One can see that
the transient times becomes arbitrarily large at the boundaries of the stability islands. Note that the transient time
is bounded from below by the dashed red curve which corresponds to the uncontrolled case according to
Eq.~(\ref{eq:transient_time}) with $\lambda=-0.062$ and $a=-0.0015$.


%




\section{Conclusion}
\label{sec:conclusion}
In conclusion, we have shown that the transient times for control of steady states by time-delayed feedback
are strongly influenced by the interplay of local and global effects. In a linear delay system the transient time scales
with an inverse power law of the control gain if the boundary of stabilization is approached. In a nonlinear system,
e.g., a Hopf normal form, global effects due to coexisting stable delay-induced orbits lead to strongly modulated
transient times as a function of the initial distance $r_0$ from the fixed point. 
These results are relevant for the optimization of time-delayed feedback control.

\begin{ack}
This work was supported by Deutsche Forschungsgemeinschaft in the framework of Sfb 555. 
\end{ack}


\begin{thebibliography}{42}
\providecommand{\natexlab}[1]{#1}
\providecommand{\url}[1]{\texttt{#1}}
\providecommand{\urlprefix}{URL }
\expandafter\ifx\csname urlstyle\endcsname\relax
  \providecommand{\doi}[1]{doi:\discretionary{}{}{}#1}\else
  \providecommand{\doi}{doi:\discretionary{}{}{}\begingroup
  \urlstyle{rm}\Url}\fi

\bibitem[{Amann et~al.(2007)Amann, Sch{\"o}ll, and Just}]{AMA05}
Amann, A., Sch{\"o}ll, E., and Just, W. (2007).
\newblock \emph{Physica~A}, 373, 191.

\bibitem[{Balanov et~al.(2006)Balanov, Beato, Janson, Engel, and
  Sch{\"o}ll}]{BAL06}
Balanov, A.G., Beato, V., Janson, N.B., Engel, H., and Sch{\"o}ll, E. (2006).
\newblock \emph{Phys.~Rev.~E}, 74, 016214.

\bibitem[{Blyuss et~al.(2008)Blyuss, Kyrychko, H{\"o}vel, and
  Sch{\"o}ll}]{BLY08}
Blyuss, K.B., Kyrychko, Y.N., H{\"o}vel, P., and Sch{\"o}ll, E. (2008).
\newblock \emph{Eur.~Phys.~J.~B}, 65, 571--576.

\bibitem[{Corless et~al.(1996)Corless, Gonnet, Hare, Jeffrey, and
  Knuth}]{COR96}
Corless, R.M., Gonnet, G.H., Hare, D.E.G., Jeffrey, D.J., and Knuth, D.E.
  (1996).
\newblock \emph{Adv. Comput. Math}, 5, 329.

\bibitem[{Dahms et~al.(2007)Dahms, H{\"o}vel, and Sch{\"o}ll}]{DAH07}
Dahms, T., H{\"o}vel, P., and Sch{\"o}ll, E. (2007).
\newblock \emph{Phys.~Rev.~E}, 76(5), 056201.

\bibitem[{Dahms et~al.(2008)Dahms, H{\"o}vel, and Sch{\"o}ll}]{DAH08b}
Dahms, T., H{\"o}vel, P., and Sch{\"o}ll, E. (2008).
\newblock \emph{Phys.~Rev.~E}, 78(5), 056213.

\bibitem[{Fiedler et~al.(2007)Fiedler, Flunkert, Georgi, H{\"o}vel, and
  Sch{\"o}ll}]{FIE07}
Fiedler, B., Flunkert, V., Georgi, M., H{\"o}vel, P., and Sch{\"o}ll, E.
  (2007).
\newblock \emph{Phys.~Rev.~Lett.}, 98, 114101.

\bibitem[{Flunkert and Sch{\"o}ll(2007)}]{FLU07}
Flunkert, V. and Sch{\"o}ll, E. (2007).
\newblock \emph{Phys. Rev. E}, 76, 066202.

\bibitem[{Fradkov(2007)}]{FRA07}
Fradkov, A.L. (2007).
\newblock \emph{{Cybernetical Physics: From Control of Chaos to Quantum
  Control}}.
\newblock Springer, Heidelberg, Germany.

\bibitem[{H{\"o}hne et~al.(2007)H{\"o}hne, Shirahama, Choe, Benner, Pyragas,
  and Just}]{HOE07a}
H{\"o}hne, K., Shirahama, H., Choe, C.U., Benner, H., Pyragas, K., and Just, W.
  (2007).
\newblock \emph{Phys.~Rev.~Lett.}, 98(21), 214102.

\bibitem[{H{\"o}vel and Sch{\"o}ll(2005)}]{HOE05}
H{\"o}vel, P. and Sch{\"o}ll, E. (2005).
\newblock \emph{Phys.~Rev.~E}, 72, 046203.

\bibitem[{Just et~al.(2004)Just, Benner, and {v. L{\"o}wenich}}]{JUS04}
Just, W., Benner, H., and {v. L{\"o}wenich}, C. (2004).
\newblock \emph{Physica~D}, 199, 33.

\bibitem[{Just et~al.(2007)Just, Fiedler, Flunkert, Georgi, H{\"o}vel, and
  Sch{\"o}ll}]{JUS07}
Just, W., Fiedler, B., Flunkert, V., Georgi, M., H{\"o}vel, P., and Sch{\"o}ll,
  E. (2007).
\newblock \emph{Phys.~Rev.~E}, 76(2), 026210.

\bibitem[{Ott et~al.(1990)Ott, Grebogi, and Yorke}]{OTT90}
Ott, E., Grebogi, C., and Yorke, J.A. (1990).
\newblock \emph{Phys.~Rev.~Lett.}, 64, 1196.

\bibitem[{Pyragas(1992)}]{PYR92}
Pyragas, K. (1992).
\newblock \emph{Phys.~Lett.~A}, 170, 421.

\bibitem[{Schikora et~al.(2006)Schikora, H{\"o}vel, W{\"u}nsche, Sch{\"o}ll,
  and Henneberger}]{SCH06a}
Schikora, S., H{\"o}vel, P., W{\"u}nsche, H.J., Sch{\"o}ll, E., and
  Henneberger, F. (2006).
\newblock \emph{Phys.~Rev.~Lett.}, 97, 213902.

\bibitem[{Schneider et~al.(2009)Schneider, Sch{\"o}ll, and Dahlem}]{SCH09c}
Schneider, F.M., Sch{\"o}ll, E., and Dahlem, M.A. (2009).
\newblock \emph{Chaos}, 19, 015110.

\bibitem[{Sch{\"o}ll(2001)}]{SCH00}
Sch{\"o}ll, E. (2001).
\newblock \emph{Nonlinear spatio-temporal dynamics and chaos in
  semiconductors}.
\newblock Cambridge University Press, Cambridge.

\bibitem[{Sch{\"o}ll(2009)}]{SCH09}
Sch{\"o}ll, E. (2009).
\newblock Pattern formation and time-delayed feedback control at the
  nano-scale.
\newblock In G.~Radons, B.~Rumpf, and H.G. Schuster (eds.), \emph{Nonlinear
  Dynamics of Nanosystems}. Wiley-VCH, Weinheim.

\bibitem[{Sch{\"o}ll et~al.(2009)Sch{\"o}ll, Hiller, H{\"o}vel, and
  Dahlem}]{SCH08}
Sch{\"o}ll, E., Hiller, G., H{\"o}vel, P., and Dahlem, M.A. (2009).
\newblock \emph{Phil.~Trans.~R.~Soc.~A}, 367, 1079--1096.

\bibitem[{Sch{\"o}ll and Schuster(2008)}]{SCH07}
Sch{\"o}ll, E. and Schuster, H.G. (eds.) (2008).
\newblock \emph{Handbook of Chaos Control}.
\newblock Wiley-VCH, Weinheim.
\newblock Second completely revised and enlarged edition.

\bibitem[{Sieber et~al.(2008)Sieber, Gonzalez-Buelga, Neild, Wagg, and
  Krauskopf}]{SIE08}
Sieber, J., Gonzalez-Buelga, A., Neild, S., Wagg, D., and Krauskopf, B. (2008).
\newblock \emph{Phys. Rev. Lett.}, 100(24), 244101.

\bibitem[{T{\'e}l and Lai(2008)}]{TEL08}
T{\'e}l, T. and Lai, Y.C. (2008).
\newblock \emph{Phys. Rep.}, 460(6), 245.

\bibitem[{Tronciu et~al.(2006)Tronciu, W{\"u}nsche, Wolfrum, and
  Radziunas}]{TRO06}
Tronciu, V.Z., W{\"u}nsche, H.J., Wolfrum, M., and Radziunas, M. (2006).
\newblock \emph{Phys.~Rev.~E}, 73, 046205.

\bibitem[{von Loewenich et~al.(2004)von Loewenich, Benner, and Just}]{LOE04}
von Loewenich, C., Benner, H., and Just, W. (2004).
\newblock \emph{Phys.~Rev.~Lett.}, 93(17), 174101--174101--4.

\bibitem[{Yanchuk and Perlikowski(2009)}]{YAN09}
Yanchuk, S. and Perlikowski, P. (2009).
\newblock \emph{Phys. Rev. E}, 79(4), 046221.

\bibitem[{Yanchuk et~al.(2006)Yanchuk, Wolfrum, H{\"o}vel, and
  Sch{\"o}ll}]{YAN06}
Yanchuk, S., Wolfrum, M., H{\"o}vel, P., and Sch{\"o}ll, E. (2006).
\newblock \emph{Phys.~Rev.~E}, 74, 026201.

\end{thebibliography}

\end{document}